\documentclass[12pt]{article}
\usepackage{amsfonts}
\usepackage{amsmath}
\usepackage{amssymb}
\usepackage{graphicx}
\usepackage{color}
\usepackage[all, knot]{xy}
\usepackage{tikz}
\usepackage{array}

\usepackage{ulem}

\usepackage[utf8]{inputenc}
\usepackage{epstopdf}
\usepackage[footnotesize]{caption}
\usepackage{amsthm}
\usepackage{enumitem}
\usepackage{mathrsfs}

\usepackage[margin=3cm]{geometry}

\def \be {\begin{equation}}
\def \ee {\end{equation}}
\def \bea {\begin{eqnarray}}
\def \eea {\end{eqnarray}}
\def \nn {\nonumber}

\def \rr {\raise.35ex\hbox{\small $\prime$}\kern-.17em{\mbox{\large $\imath$}}}

\def \dels {\partial\kern-.6em /\kern.1em}
\def \As {{A\kern-.5em / \kern.5em}}
\def \Ds {D\kern-.7em / \kern.5em}

\def \ks {k\kern-.5em /}
\def \ls {l\kern-.5em /}

\newcommand{\ci}[1]{}

\newcommand{\ba}{\begin{eqnarray}}
\newcommand{\ea}{\end{eqnarray}}
\newcommand{\bal}{\begin{align}}
\newcommand{\eal}{\end{align}}
\newcommand{\bay}[1]{\left(\begin{array}{#1}}
\newcommand{\eay}{\end{array}\right)}

\newcommand{\hide}[1]{}

\newlist{axioms}{enumerate}{2}
\setlist[axioms,1]{label=\textbf{A\arabic{axiomsi}.}, ref=A\arabic{axiomsi}}
\setlist[axioms,2]{label=\textbf{A\arabic{axiomsi}\rlap{\myEnumCounter{axiomsii}}.},%
                   ref=A\arabic{axiomsi}\myEnumCounter{axiomsii},%
                   align=parleft,%
                   leftmargin=0em,%
                   itemsep=1.4ex,%
                   before={\stepcounter{axiomsi}}}

  \usetikzlibrary{decorations.markings}

\begin{document}
\begin{titlepage}

\begin{center}

\textbf{\LARGE
Relation between Commutative and Non-Commutative Descriptions of D-branes in Large R-R Field Background 
\vskip.3cm
}
\vskip .5in
{\large
Chen-Te Ma$^{a}$ \footnote{e-mail address: yefgst@gmail.com}
\\
\vskip 1mm
}
{\sl
$^a$
Department of Physics, Great Bay University, Dongguan, Guangdong 52300, China. 
}
\\
\vskip 1mm
\vspace{40pt}
\end{center}

\begin{abstract}
\noindent
We derive the Seiberg-Witten map to first order in the non-commutativity parameter for D-branes in the presence of a large R-R background field. 
This result enables a systematic investigation of the commutative formulation of the corresponding Lagrangian. 
In the SU($N$) sector, the map introduces a non-local operator. In contrast, in the U(1) sector, this non-locality can be removed. 
This contrast suggests that the essential source of non-local behavior lies in the non-Abelian degrees of freedom. 
The commutative description obtained here offers further insight into both the Dirac-Born-Infeld structure and its possible extensions to the dynamics of M5-branes.
\end{abstract}
\end{titlepage}

\section{Introduction}
\label{sec:1}
\noindent 
The modern notion of gauge theory originates from the formulation of Maxwell’s equations, which describe the electromagnetic interaction. Electromagnetism can be understood as a gauge theory based on a U(1) symmetry group. Yang–Mills (YM) theory extends this framework by promoting the Abelian gauge group to a non-Abelian one, enabling the description of elementary particle interactions. 
The use of group-theoretic language not only unifies the electromagnetic and weak forces but also provides the foundation for describing the strong interaction. 
Consequently, the development of Yang–Mills theory and the pursuit of underlying fundamental symmetries have become central pillars in our understanding of the microscopic structure of nature.
\\

\noindent 
String theory offers a natural extension of Yang–Mills (YM) theory through its underlying symmetry structure. 
In this framework, fundamental objects are one-dimensional strings—either closed loops or open segments with endpoints. 
Extended objects known as Dirichlet branes (D-branes) serve as hypersurfaces on which open strings can end, subject to Dirichlet boundary conditions. 
The massless modes of open strings terminating on a single D-brane give rise to a U(1) gauge field. 
When $N$ D-branes coincide, additional string modes stretching between different branes enlarge the gauge symmetry to U($N$) \cite{Tseytlin:1997csa}. 
Thus, string theory provides a natural setting for realizing and studying non-Abelian gauge symmetries \cite{Tseytlin:1997csa}.
A D$p$-brane denotes a D-brane extended along $p$ spatial dimensions. 
The low-energy effective dynamics of D$p$-branes are governed by the Dirac–Born–Infeld (DBI) action, which generalizes Maxwell theory by incorporating an infinite series of higher-derivative corrections \cite{Abouelsaood:1986gd,Callan:1986bc}.
\\

\noindent 
Another significant insight into gauge theory emerges from the non-commutative structure induced by quantizing open strings in the presence of a Neveu–Schwarz–Neveu–Schwarz (NS–NS) background field \cite{Chu:1998qz}. 
In this context, the Seiberg–Witten (SW) map provides an explicit correspondence between {\it commutative} and {\it non-commutative} gauge theories \cite{Seiberg:1999vs,Cornalba:1999ah,Okawa:1999cm,Asakawa:1999cu,Ishibashi:1999vi}. 
Because the SW map acts as a field redefinition, the equivalence between the two descriptions requires that the map commute with gauge transformations \cite{Seiberg:1999vs}. 
Although commutative and non-commutative formulations are physically equivalent, the non-commutative gauge theory conveniently encodes higher-derivative corrections through the Moyal product, already at leading order in the non-commutativity parameter (the inverse of the NS–NS background field) \cite{Seiberg:1999vs}. 
This makes the non-commutative description particularly advantageous for exploring the symmetry structure of higher-derivative terms \cite{Seiberg:1999vs}.
\\

\noindent 
D-branes act as sources for electric and magnetic Ramond–Ramond (R–R) fields. 
A Lagrangian formulation of D-branes in an R–R background \cite{Cornalba:2002cu} is therefore essential for elucidating the network of dualities that organize M-theory. 
M-theory itself is proposed as a unifying framework for all fundamental interactions, reducing at low energies to eleven-dimensional supergravity—much as string theory reduces to ten-dimensional supergravity in its low-energy limit. 
In eleven-dimensional supergravity, the membrane (M2-brane) couples electrically to the three-form $C$-field, while its magnetic dual is the M5-brane.
\\

\noindent 
A distinctive feature of M2-branes is that their entropy scaling differs from that of D-branes. 
This indicates that an ordinary Lie algebra does not govern the gauge sector on multiple M2-branes but rather by a higher algebraic structure—a Lie 3-algebra. 
The first concrete realization of this idea is the Bagger–Lambert–Gustavsson (BLG) model, which describes two coincident M2-branes \cite{Bagger:2006sk,Gustavsson:2007vu,Bagger:2007jr,Bagger:2007vi}. 
Extending Lie 3-algebra constructions to define a finite number of M2-branes generally requires relaxing the condition of a positive-definite metric \cite{Ho:2008ei}. 
In contrast, when one considers an infinite-dimensional Lie 3-algebra, an infinite stack of M2-branes becomes equivalent to a single M5-brane defined on a non-commutative space \cite{Ho:2008nn,Ho:2008ve}. 
The resulting gauge theory on the M5-brane is a self-dual gauge theory \cite{Ho:2008nn,Ho:2008ve}, in stark contrast with the Yang–Mills structure familiar from D-branes—precisely as suggested by the distinct entropy scaling behavior.
\\

\noindent 
The dynamics of a single M5-brane in a large $C$-field background are governed by the Nambu–Poisson (NP) bracket, which corresponds to an infinite-dimensional Lie 3-algebra \cite{Ho:2008nn,Ho:2008ve}. 
The resulting formulation is known as the NP M5-brane theory \cite{Ho:2008nn,Ho:2008ve,Ho:2012dn}. 
In this framework, the gauge symmetry is generated by volume-preserving diffeomorphisms (VPDs), themselves induced by the NP bracket \cite{Ho:2008nn,Ho:2008ve}. 
The NP M5-brane theory can be interpreted as the large-$C$-field limit of the complete M5-brane theory, a limit in which Lorentz symmetry is explicitly broken \cite{Ho:2008nn,Ho:2008ve}. 
Accordingly, the worldvolume directions naturally separate into those parallel and those transverse to the background flux \cite{Ho:2008nn,Ho:2008ve}.
\\

\noindent 
Upon compactification, different choices of compact direction lead to distinct effective theories. 
Compactifying a direction parallel to the background yields the non-commutative D4-brane with a large NS–NS $B$-field \cite{Ho:2008ve}, while compactifying a direction orthogonal to the background produces the non-commutative D4-brane with a large R–R $C$-field \cite{Ho:2011yr,Ma:2012hc}. 
In the latter case, the VPD gauge symmetry survives and plays a novel role within the D-brane context \cite{Ho:2011yr}. 
The Lagrangian formulation of a D$p$-brane in a large R–R ($p-1$)-form background can be systematically constructed by combining partial Lorentz symmetry, U(1) gauge symmetry, VPD gauge symmetry, and consistent dimensional reduction \cite{Ho:2013paa}. 
In particular, for $p=3$, the NS–NS D3-brane is related by electromagnetic duality to the R–R D3-brane, providing nontrivial support for the R–R D-brane construction \cite{Ho:2013opa,Ho:2015mfa}. 
Extending the single R–R D-brane theory to the case of multiple branes is achieved by replacing partial derivatives with covariant derivatives \cite{Ma:2023hgi}. 
This preserves the VPD symmetry while enlarging the U(1) gauge symmetry to U($N$) \cite{Ma:2023hgi}. 
At leading order, the resulting non-commutativity parameter reproduces the structure of U($N$) Yang–Mills theory \cite{Ma:2023hgi}. 
Moreover, the R–R D4-branes theory is expected to uplift, via dimensional reduction along a direction transverse to the large $C$-field, to a corresponding description of multiple M5-branes. 
One promising approach introduces a five-dimensional non-dynamical one-form gauge field that couples to the six-dimensional self-dual two-form gauge field \cite{Chu:2011fd,Chu:2012um}, suggesting that the non-Abelian sector of the M5-branes exhibits behavior qualitatively {\it distinct} from that of the Abelian case.
\\

\begin{figure}
\begin{center}
\includegraphics[width=1.\textwidth]{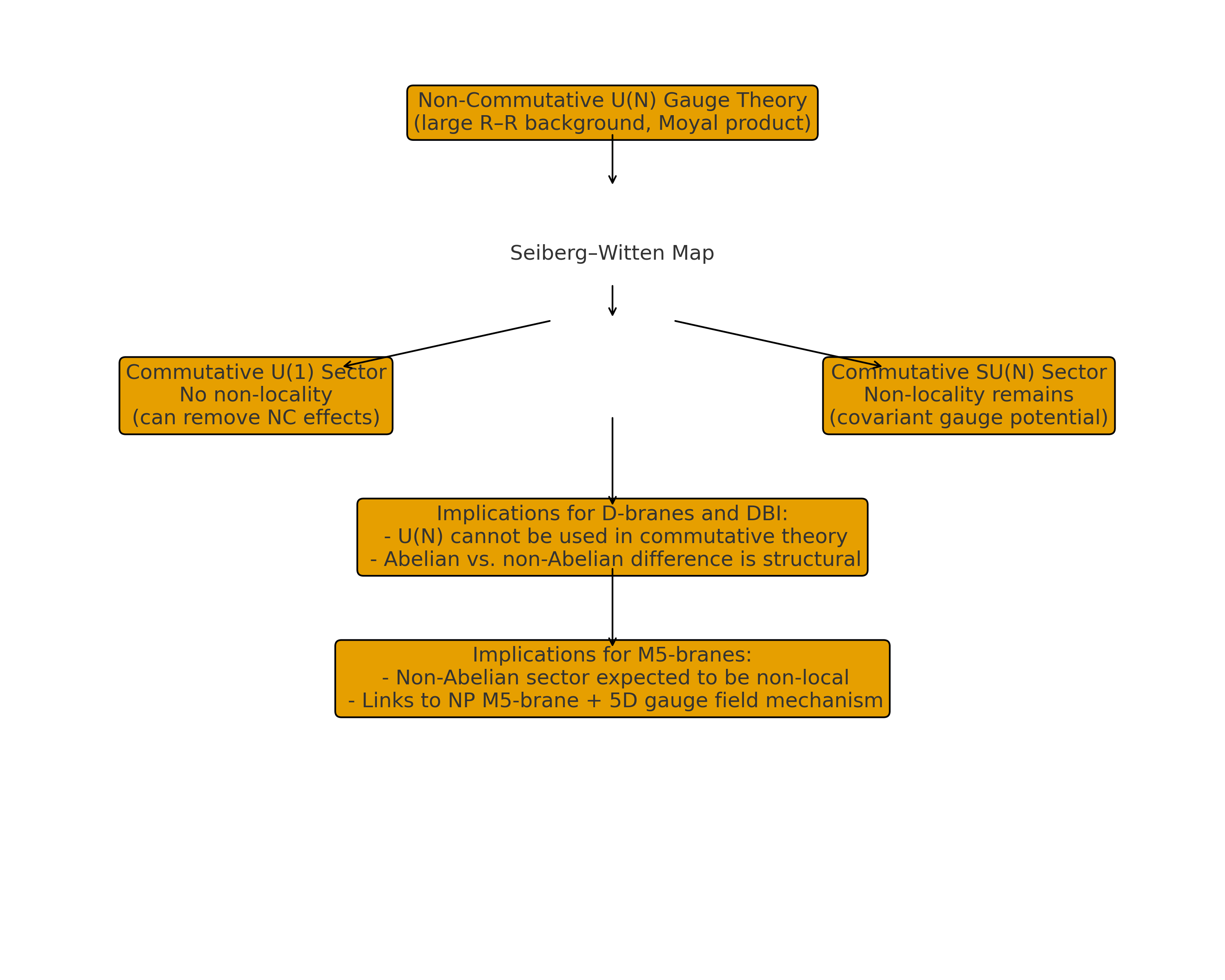}
\end{center}
\caption{The result's summarization and implication. }
\label{summary.png}
\end{figure}

\noindent
In this paper, we investigate the distinction between the Abelian and non-Abelian sectors through the commutative description of R–R D-branes.
In the non-Abelian sector, non-locality arises explicitly through the appearance of the covariant gauge potential.
In contrast, in the Abelian sector, this non-locality disappears.
This contrast indicates that the non-local features are intrinsically tied to the non-Abelian structure.
The main results of our work are presented in Fig. \ref{summary.png}.
Our main results are summarized as follows:
\begin{itemize}
\item We first examine the gauge transformations of R–R D-branes and demonstrate that a naive extension to curved backgrounds fails to produce a closed gauge algebra.
For this reason, we restrict our analysis to flat backgrounds.
\item We generalize the solution of the Seiberg–Witten (SW) map from the U(1) case \cite{Ma:2020msx} to U($N$).
The resulting structure reveals that the commutative Lagrangian depends on the SU($N$) covariant gauge potential, not solely on the covariant field strength.
This finding rules out a naive extension of the DBI action obtained simply by promoting the U(1) gauge group to U($N$) in the commutative description.
\item Although the Abelian and non-Abelian sectors exhibit different commutative descriptions, they become unified within the U($N$) framework of the non-commutative theory.
This observation suggests that higher-derivative corrections for M5-branes may be more naturally understood from the non-commutative perspective.
\end{itemize}

\noindent 
The structure of the paper is as follows. 
In Sec.~\ref{sec:2}, we discuss the gauge transformations. 
In Sec.~\ref{sec:3}, we present the solution of the SW map for R–R D-branes. 
In Sec.~\ref{sec:4}, we analyze the commutative Lagrangian and discuss its implications for the DBI structure and for the dynamics of the M5-branes. 
We conclude in Sec.~\ref{sec:5} with a summary and discussion of our results.


\section{Gauge Transformation}
\label{sec:2}
\noindent
We first show the gauge transformation of the R-R D$p$-branes \cite{Ma:2023hgi}:
\bea
\hat{\delta}_{\hat{\Lambda}}\hat{b}^{\dot{\mu}}&=&\hat{\kappa}^{\dot{\mu}}
+i\lbrack\hat{\lambda}, \hat{b}^{\dot\mu}\rbrack
+\theta\hat{\kappa}^{\dot{\nu}}\partial_{\dot{\nu}}\hat{b}^{\dot{\mu}};
\nn\\
\hat{\delta}_{\hat{\Lambda}}\hat{B}_{\alpha}{}^{\dot{\mu}}&=&\partial_{\alpha}\hat{\kappa}^{\dot{\mu}}
+i\lbrack\hat{\lambda}, \hat{B}_{\alpha}{}^{\dot\mu}\rbrack
+\theta\big(\hat{\kappa}^{\dot{\nu}}\partial_{\dot{\nu}}\hat{B}_{\alpha}{}^{\dot{\mu}}-(\partial_{\dot{\nu}}\hat{\kappa}^{\dot{\mu}})\hat{B}_{\alpha}{}^{\dot{\nu}}\big);
\nn\\
\hat{\delta}_{\hat{\Lambda}}\hat{a}_{\alpha}&=&\partial_{\alpha}{\hat{\lambda}}
+i\lbrack\hat{\lambda}, \hat{a}_{\alpha}\rbrack
+\theta(\hat{\kappa}^{\dot{\nu}}\partial_{\dot{\nu}}\hat{a}_{\alpha}
+\hat{a}_{\dot{\nu}}\partial_{\alpha}\hat{\kappa}^{\dot{\nu}});
\nn\\
\hat{\delta}_{\hat{\Lambda}}\hat{a}_{\dot{\mu}}&=&\partial_{\dot{\mu}}{\hat{\lambda}}
+i\lbrack\hat{\lambda}, \hat{a}_{\dot\mu}\rbrack
+\theta(\hat{\kappa}^{\dot{\nu}}\partial_{\dot{\nu}}\hat{a}_{\dot{\mu}}
+\hat{a}_{\dot{\nu}}\partial_{\dot{\mu}}\hat{\kappa}^{\dot{\nu}}),
\eea
where $\hat{B}_{\alpha}{}^{\dot{\mu}}$ satisfies that \cite{Ma:2023hgi}
\bea
&&\hat{V}_{\dot{\mu}}{}^{\dot{\nu}}(\lbrack\hat{D}^{\alpha}, \hat{b}_{\dot{\nu}}\rbrack-\hat{V}^{\dot{\rho}}{}_{\dot{\nu}}\hat{B}^{\alpha}{}_{\dot{\rho}})
+\epsilon^{\alpha\beta}\hat{F}_{\beta\dot{\mu}}
+\theta\epsilon^{\alpha\beta}\hat{F}_{\dot{\mu}\dot{\nu}}\hat{B}_{\beta}{}^{\dot{\nu}}=0.
\eea
The $\hat{V}_{\dot\nu}{}^{\dot\mu}$ is as \cite{Ma:2023hgi}
\bea
\hat{V}_{\dot\mu}{}^{\dot\mu}=\delta_{\dot\nu}{}^{\dot\mu}
+\theta\lbrack\hat{D}_{\dot\nu}, \hat{b}^{\dot\mu}\rbrack,
\eea
where
\bea
\lbrack\hat{D}_{\dot\nu}, \hat{b}^{\dot\mu}\rbrack=\partial_{\dot\nu}\hat{b}^{\dot\mu}
-i\lbrack\hat{a}_{\dot\nu}, \hat{b}^{\dot\mu}\rbrack.
\eea
The range of indices are $\alpha=0, 1$ and $\dot\mu=2, 3,\cdots, p$.
The gauge parameter $\hat{\kappa}$ is a divergenless parameter,
\bea
\partial_{\dot\mu}\hat{\kappa}^{\dot{\mu}}=0,
\eea
generates the VPD symmetry, and the $\hat{\lambda}$ generates the U($N$) symmetry with the adjoint representation \cite{Ma:2023hgi}.
Because the VPD gauge parameter controls the large field background, it only resides in the U(1) sector \cite{Ma:2023hgi}.
The $\theta$ is a dimensionless non-commutativity parameter that serves as a perturbation parameter in the R-R D-branes.
The $\hat{b}^{\dot\mu}$ is not a dynamical field, and the field strength does not have a time derivative term \cite{Ma:2023hgi}.
The $\hat{B}_{\alpha}{}^{\dot\mu}$ is a variable collecting all fields of the R-R D-branes \cite{Ma:2023hgi}.
Hence, all physical degrees of freedom are given by the one-form dynamical fields, $\hat{a}_{\alpha}$ and $\hat{a}_{\dot\mu}$ \cite{Ma:2023hgi}.
When we replace the covariant derivative with the partial derivative, the multiple D-branes are reduced to a single D-brane \cite{Ma:2023hgi}.
The naive question is whether we can generalize from a flat background to a general curved background by simply modifying the covariant derivative.
Let us examine the following gauge transformation
\bea
\hat{\delta}_{\hat{\Lambda}}\hat{b}^{\dot{\mu}}=\hat{\kappa}^{\dot{\mu}}
+i\lbrack\hat{\lambda}, \hat{b}^{\dot\mu}\rbrack
+\theta\hat{\kappa}^{\dot{\nu}}\hat{\nabla}_{\dot{\nu}}\hat{b}^{\dot{\mu}},
\eea
where
\bea
\hat{\nabla}_{\dot\mu}\hat{\kappa}^{\dot\mu}=0.
\eea
In general, the gauge transformation is not closed
\bea
\lbrack\hat{\delta}_{\hat{\Lambda}_2}, \hat{\delta}_{\hat{\Lambda}_1}\rbrack\hat{b}^{\dot\mu}
=\hat{\tilde{\kappa}}^{\dot\mu}
+i\lbrack\hat{\tilde{\lambda}}, \hat{b}^{\dot\mu}\rbrack
+\theta\hat{\tilde{\kappa}}^{\dot\rho}\hat{\nabla}_{\dot\rho}\hat{b}^{\dot\mu}
+\hat{\kappa}_1^{\dot\rho}\hat{\kappa}_2^{\dot\sigma}\lbrack\hat{\nabla}_{\dot\rho}, \hat{\nabla}_{\dot\sigma}\rbrack\hat{b}^{\dot\mu},
\eea
where
\bea
\hat{\tilde{\kappa}}^{\dot\mu}=\theta(\hat{\kappa}_1^{\dot\rho}\hat{\nabla}_{\dot\rho}\hat{\kappa}_2^{\dot\mu}-\hat{\kappa}_2^{\dot\rho}\hat{\nabla}_{\dot\rho}\hat{\kappa}_1^{\dot\mu}); \
\hat{\tilde{\lambda}}=i\lbrack\hat{\lambda}_1, \hat{\lambda}_2\rbrack
+(\hat{\kappa}_1^{\dot\rho}\hat{\nabla}_{\dot\rho}\hat{\lambda}_2
-\hat{\kappa}_2^{\dot\rho}\hat{\nabla}_{\dot\rho}\hat{\lambda}_2).
\eea
Only when the commutator of the covariant derivative vanishes can we have a closed gauge transformation.
It suggests that the naive generalization, or one similar to the non-Abelization, cannot be applied to a generic curved background.
Hence, in this paper, we will only focus on the flat background.

\section{SW Map}
\label{sec:3}
\noindent
In this section, we aim to extend the solution of the SW map from the U(1) gauge group to the U($N$) gauge group.
When the SW map commutes with the gauge transformation, it equivalently satisfies the following relations:
\bea
&&
\hat{b}^{\dot{\mu}}(b+\delta_{\Lambda}b)-\hat{b}^{\dot{\mu}}(b)=\hat{\delta}_{\hat{\Lambda}}\hat{b}^{\dot{\mu}};
\nn\\
&&
\hat{a}_{\alpha}(a+\delta_{\Lambda}a, b+\delta_{\Lambda}b)-\hat{a}_{\alpha}(a, b)=\hat{\delta}_{\hat{\Lambda}}\hat{a}_{\alpha}; \
\hat{a}_{\dot\mu}(a+\delta_{\Lambda}a, b+\delta_{\Lambda}b)-\hat{a}_{\dot\mu}(a, b)=\hat{\delta}_{\hat{\Lambda}}\hat{a}_{\dot\mu},
\nn\\
\eea
where the gauge transformation for the commutative description is:
\bea
\delta_{\Lambda}b^{\dot\mu}=\kappa^{\dot\mu}+i\lbrack\lambda, b^{\dot\mu}\rbrack; \
\delta_{\Lambda}a_{\alpha}=\partial_{\alpha}\lambda+i\lbrack\lambda, a_{\alpha}\rbrack; \
\delta_{\Lambda}a_{\dot\mu}=\partial_{\dot\mu}\lambda+i\lbrack\lambda, a_{\dot\mu}\rbrack.
\eea
Ref. \cite{Ma:2020msx} can deduce the solution of the SW map for the single D-brane:
\bea
\hat{b}^{\dot{\mu}, \mathrm{U(1)}}(b)&=&b^{\dot{\mu}, \mathrm{U(1)}}+\theta\bigg(\frac{1}{2}b^{\dot{\nu}, \mathrm{U(1)}}\partial_{\dot{\nu}}b^{\dot{\mu}, \mathrm{U(1)}}
+\frac{1}{2}b^{\dot{\mu}, \mathrm{U(1)}}\partial_{\dot{\nu}}b^{\dot{\nu}, \mathrm{U(1)}}\bigg)+{\cal O}(\theta^2);
\nn\\
\hat{a}^{\mathrm{U(1)}}_{\alpha}(a, b)&=&a^{\mathrm{U(1)}}_{\alpha}
+\theta(b^{\dot{\rho}, \mathrm{U(1)}}\partial_{\dot{\rho}}a^{\mathrm{U(1)}}_{\alpha}+a^{\mathrm{U(1)}}_{\dot{\rho}}\partial_{\alpha}b^{\dot{\rho}, \mathrm{U(1)}})+{\cal O}(\theta^2);
\nn\\
\hat{a}_{\dot\mu}^{\mathrm{U(1)}}(a, b)&=&a^{\mathrm{U(1)}}_{\dot\mu}
+\theta(b^{\dot{\rho}, \mathrm{U(1)}}\partial_{\dot{\rho}}a_{\dot\mu}^{\mathrm{U(1)}}+a_{\dot{\rho}}^{\mathrm{U(1)}}\partial_{\dot\mu}b^{\dot{\rho}, \mathrm{U(1)}})+{\cal O}(\theta^2);
\nn\\
\hat{\kappa}^{\dot{\mu}}&=&\kappa^{\dot{\mu}}
+\theta\bigg(\frac{1}{2}b^{\dot{\nu}, \mathrm{U(1)}}\partial_{\dot{\nu}}\kappa^{\dot{\mu}}
+\frac{1}{2}(\partial_{\dot{\nu}}b^{\dot{\nu}, \mathrm{U(1)}})\kappa^{\dot{\mu}}
-\frac{1}{2}(\partial_{\dot{\nu}}b^{\dot{\mu}, \mathrm{U(1)}})\kappa^{\dot{\nu}}\bigg)
+{\cal O}(\theta^2);
\nn\\
\hat{\lambda}^{\mathrm{U(1)}}&=&\lambda^{\mathrm{U(1)}}+\theta b^{\dot{\rho}, \mathrm{U(1)}}\partial_{\dot{\rho}}\lambda^{\mathrm{U(1)}}+{\cal O}(\theta^2).
\eea
Because the VPD gauge parameter is not introduced in the SU($N$) part, and $\hat{b}^{\dot\mu, \mathrm{SU(N)}}$ becomes covariant, the non-Abelian part behaves differently from the Abelian part.
\\

\noindent
Any U($N$) field can be decomposed into its U(1) and SU($N$) components.
We can first find the solution of the SU($N$) sector and then combine it with the U(1) part to get the solution for the U($N$) fields.
The solution of the SW map for the SU($N$) sector is:
\bea
\hat{b}^{\dot\mu, \mathrm{SU(N)}}(b)&=&
b^{\dot\mu, \mathrm{SU(N)}}
+\theta b^{\dot\rho, \mathrm{U(1)}}\partial_{\dot\rho}b^{\dot\mu, \mathrm{SU(N)}}+{\cal O}(\theta^2);
\nn\\
\hat{a}^{\mathrm{SU(N)}}_{\alpha}(a, b)&=&a^{\mathrm{SU(N)}}_{\alpha}+\theta(b^{\dot{\rho}, \mathrm{U(1)}}\partial_{\dot{\rho}}a_{\alpha}^{\mathrm{SU(N)}}+a^{\mathrm{SU(N)}}_{\dot{\rho}}\partial_{\alpha}b^{\dot{\rho}, \mathrm{U(1)}})+{\cal O}(\theta^2);
\nn\\
\hat{a}^{\mathrm{SU(N)}}_{\dot\mu}(a, b)&=&a^{\mathrm{SU(N)}}_{\dot\mu}+\theta(b^{\dot{\rho}, \mathrm{U(1)}}\partial_{\dot{\rho}}a_{\dot\mu}^{\mathrm{SU(N)}}+a^{\mathrm{SU(N)}}_{\dot{\rho}}\partial_{\dot\mu}b^{\dot{\rho}, \mathrm{U(1)}})+{\cal O}(\theta^2).
\eea
We now combine it with the U(1) case to show the following U($N$) solution:
\bea
\hat{b}^{\dot{\mu}}(b)&=&b^{\dot{\mu}}+\theta\bigg(
b^{\dot{\nu}, \mathrm{U(1)}}\partial_{\dot{\nu}}b^{\dot{\mu}, \mathrm{SU(N)}}
+\frac{1}{2}b^{\dot{\nu}, \mathrm{U(1)}}\partial_{\dot{\nu}}b^{\dot{\mu}, \mathrm{U(1)}}
+\frac{1}{2}b^{\dot{\mu}, \mathrm{U(1)}}\partial_{\dot{\nu}}b^{\dot{\nu}, \mathrm{U(1)}}\bigg)+{\cal O}(\theta^2);
\nn\\
\hat{a}_{\alpha}(a, b)&=&a_{\alpha}
+\theta(b^{\dot{\rho}, \mathrm{U(1)}}\partial_{\dot{\rho}}a_{\alpha}+a_{\dot{\rho}}\partial_{\alpha}b^{\dot{\rho}, \mathrm{U(1)}})+{\cal O}(\theta^2);
\nn\\
\hat{a}_{\dot\mu}(a, b)&=&a_{\dot\mu}
+\theta(b^{\dot{\rho}, \mathrm{U(1)}}\partial_{\dot{\rho}}a_{\dot\mu}+a_{\dot{\rho}}\partial_{\dot\mu}b^{\dot{\rho}, \mathrm{U(1)}})+{\cal O}(\theta^2);
\nn\\
\hat{\kappa}^{\dot{\mu}}&=&\kappa^{\dot{\mu}}
+\theta\bigg(\frac{1}{2}b^{\dot{\nu}, \mathrm{U(1)}}\partial_{\dot{\nu}}\kappa^{\dot{\mu}}
+\frac{1}{2}(\partial_{\dot{\nu}}b^{\dot{\nu}, \mathrm{U(1)}})\kappa^{\dot{\mu}}
-\frac{1}{2}(\partial_{\dot{\nu}}b^{\dot{\mu}, \mathrm{U(1)}})\kappa^{\dot{\nu}}\bigg)
+{\cal O}(\theta^2);
\nn\\
\hat{\lambda}&=&\lambda+\theta b^{\dot{\rho}, \mathrm{U(1)}}\partial_{\dot{\rho}}\lambda+{\cal O}(\theta^2),
\eea
In the commutative framework, the U(1) and SU($N$) fields cannot be combined into a U($N$) field, unlike in the non-commutative setting.
Hence, the solution structure already shows the difference between the Abelian and non-Abelian sectors in the commutative description.
We now show the non-Abelian field strengths for the commutative descriptions, which help show the Lagrangian in the commutative description:
\bea
&&
\hat{H}_{\dot1\dot2\dot3}
\nn\\
&=&H_{\dot1\dot2\dot3}
\nn\\
&&
+\theta\bigg((\partial_{\dot\mu}b^{\dot\rho, \mathrm{U(1)}})\lbrack D_{\dot\rho}, b^{\dot\mu, \mathrm{SU(N)}}\rbrack
+b^{\dot\rho, \mathrm{U(1)}}\partial_{\dot\rho}(\lbrack D_{\dot\mu}, b^{\dot\mu, \mathrm{SU(N)}}\rbrack)
\nn\\
&&
+\frac{1}{2}(\partial_{\dot\mu}b^{\dot\rho, \mathrm{U(1)}})(\partial_{\dot\rho}b^{\dot\mu, \mathrm{U(1)}})
+b^{\dot\rho, \mathrm{U(1)}}(\partial_{\dot\mu}\partial_{\dot\rho}b^{\dot\mu, \mathrm{U(1)}})
+\frac{1}{2}(\partial_{\dot\mu}b^{\dot\mu, \mathrm{U(1)}})(\partial_{\dot\rho}b^{\dot\rho, \mathrm{U(1)}})
\bigg)
+{\cal O}(\theta^2);
\nn\\
&&
\hat{F}_{\alpha\dot\mu}
\nn\\
&=&F_{\alpha\dot\mu}
\nn\\
&&
+\theta\big(
(\partial_{\alpha}b^{\dot\rho, \mathrm{U(1)}})F_{\dot\rho\dot\mu}
+b^{\dot\rho, \mathrm{U(1)}}(\partial_{\dot\rho}F_{\alpha\dot\mu})
+(\partial_{\dot\mu}b^{\dot\rho, \mathrm{U(1)}})F_{\alpha\dot\rho}\big)
+{\cal O}(\theta^2);
\nn\\
&&
\hat{F}_{\dot\mu\dot\nu}
\nn\\
&=&F_{\dot\mu\dot\nu}
\nn\\
&&
+\theta\big(
(\partial_{\dot\mu}b^{\dot\rho, \mathrm{U(1)}})F_{\dot\rho\dot\nu}
+b^{\dot\rho, \mathrm{U(1)}}(\partial_{\dot\rho}F_{\dot\mu\dot\nu})
+(\partial_{\dot\nu}b^{\dot\rho, \mathrm{U(1)}})F_{\dot\mu\dot\rho}\big)
+{\cal O}(\theta^2);
\nn\\
&&
\hat{F}_{\alpha\beta}
\nn\\
&=&F_{\alpha\beta}
\nn\\
&&
+\theta\big(
(\partial_{\alpha}b^{\dot\rho, \mathrm{U(1)}})F_{\dot\rho\beta}
+b^{\dot\rho, \mathrm{U(1)}}(\partial_{\dot\rho}F_{\alpha\beta})
+(\partial_{\dot\beta}b^{\dot\rho, \mathrm{U(1)}})F_{\alpha\dot\rho}\big)
+{\cal O}(\theta^2).
\eea

\section{Commutative Description}
\label{sec:4}
\noindent
We begin from the Lagrangian for the gauge sector of the R-R D4-branes \cite{Ma:2023hgi}
\bea
S_{\mathrm{RR4}}=\frac{T_4}{\theta^2}\int d^2xd^3\dot{y}\ \mathrm{Str}\bigg\lbrack\bigg(-\frac{1}{2}\hat{{\cal H}}_{\dot{1}\dot{2}\dot{3}}\hat{{\cal H}}^{\dot{1}\dot{2}\dot{3}}
-\frac{1}{4}\hat{{\cal F}}_{\dot{\nu}\dot{\rho}}\hat{{\cal F}}^{\dot{\nu}\dot{\rho}}
+\frac{1}{2}\hat{{\cal F}}_{\beta\dot{\mu}}\hat{{\cal F}}^{\beta\dot{\mu}}
+\frac{1}{2\theta}\epsilon^{\alpha\beta}\hat{{\cal F}}_{\alpha\beta}\bigg)\bigg\rbrack,
\nn\\
\eea
where $T_4$ is the tension of the D4-branes.
The expression of the field strengths is \cite{Ma:2023hgi}:
\bea
\hat{{\cal H}}_{\dot{1}\dot{2}\dot{3}}&\equiv&
\lbrack\hat{D}_{\dot{\mu}}, \hat{b}^{\dot{\mu}}\rbrack
+\frac{\theta}{2}(\lbrack\hat{D}_{\dot{\nu}}, \hat{b}^{\dot{\nu}}\rbrack\lbrack\hat{D}_{\dot{\rho}}, \hat{b}^{\dot{\rho}}\rbrack
-\lbrack\hat{D}_{\dot{\nu}}, \hat{b}^{\dot{\rho}}\rbrack\lbrack\hat{D}_{\dot{\rho}}, \hat{b}^{\dot{\nu}}\rbrack)
+\theta^2\{\hat{b}^{\dot{1}}, \hat{b}^{\dot{2}}, \hat{b}^{\dot{3}}\},
\nn\\
\hat{{\cal F}}_{\dot{\mu}\dot{\nu}}&\equiv& \hat{F}_{\dot{\mu}\dot{\nu}}
+\theta(\lbrack\hat{D}_{\dot{\sigma}}, \hat{b}^{\dot{\sigma}}\rbrack\hat{F}_{\dot{\mu}\dot{\nu}}
-\lbrack\hat{D}_{\dot{\mu}}, \hat{b}^{\dot{\sigma}}\rbrack\hat{F}_{\dot{\sigma}\dot{\nu}}
-\lbrack\hat{D}_{\dot{\nu}}, \hat{b}^{\dot{\sigma}}\rbrack\hat{F}_{\dot{\mu}\dot{\sigma}}),
\nn\\
\hat{{\cal F}}_{\alpha\dot{\mu}}&\equiv& (\hat{V}^{-1})_{\dot{\mu}}{}^{\dot{\nu}}(\hat{F}_{\alpha\dot{\nu}}
+\theta\hat{F}_{\dot{\nu}\dot{\sigma}}\hat{B}_{\alpha}{}^{\dot{\sigma}}),
\nn\\
\hat{{\cal F}}_{\alpha\beta}&=&\hat{F}_{\alpha\beta}
+\theta(-\hat{F}_{\alpha\dot{\mu}}\hat{B}_{\beta}{}^{\dot{\mu}}
-\hat{F}_{\dot{\mu}\beta}\hat{B}_{\alpha}{}^{\dot{\mu}})
+\frac{\theta^2}{2}\hat{F}_{\dot{\mu}\dot{\nu}}(\hat{B}_{\alpha}{}^{\dot{\mu}}\hat{B}_{\beta}{}^{\dot{\nu}}+\hat{B}_{\beta}{}^{\dot\nu}\hat{B}_{\alpha}{}^{\dot\mu})
,
\eea
where
\bea
\{ \hat{{\cal O}}_1, \hat{{\cal O}}_2, \hat{{\cal O}}_3\}\equiv\epsilon^{\dot{\mu}\dot{\nu}\dot{\rho}}\lbrack\hat{D}_{\dot{\mu}}, \hat{{\cal O}}_1\rbrack\lbrack\hat{D}_{\dot{\nu}}, \hat{{\cal O}}_2\rbrack\lbrack\hat{D}_{\dot{\rho}}, \hat{{\cal O}}_3\rbrack.
\eea
We use the symmetrized trace \cite{Tseytlin:1997csa,Ma:2023hgi}
\bea
\mathrm{Str}({\cal O}_1{\cal O}_2\cdots{\cal O}_n)&\equiv&
\mathrm{Tr}\big(\mathrm{Sym}({\cal O}_1{\cal O}_2\cdots{\cal O}_n)\big);
\nn\\
\mathrm{Sym}({\cal O}_1{\cal O}_2\cdots{\cal O}_n)&\equiv&\frac{1}{n!}({\cal O}_1{\cal O}_2\cdots {\cal O}_n+ \mathrm{all\ permutations}).
\eea
The SU($N$) generator satisfies the following properties:
\bea
\mathrm{Tr}(T^a)=0; \qquad \mathrm{Tr}(T^aT^b)=N\delta^{ab}; \qquad \mathrm{Tr}(T^aT^bT^c)=\frac{1}{2}iNf^{abc}.
\eea
To write the Lagrangian in the commutative description from the first-order correction of the non-commutativity parameter, we first show the perturbation result and also the commutative description of the covariant field strengths.
We will concentrate on the first-order term of the Lagrangian, as the leading-order term is the U($N$) Yang-Mills term, which was demonstrated in Ref. \cite{Ma:2023hgi}.
We will also discuss the implications of the DBI form and the M5-branes from the results of the commutative description.

\subsection{Perturbation}
\noindent
We show the following useful perturbation solution that we need for writing the covariant field strengths and Lagrangian
\bea
&&
\hat{B}_{\alpha}{}^{\dot\mu}
\nn\\
&=&\epsilon_{\alpha\beta}\hat{F}^{\beta\dot\mu}+\lbrack\hat{D}_{\alpha}, \hat{b}^{\dot\mu}\rbrack
\nn\\
&&
+\theta(-\lbrack\hat{D}^{\dot\rho}, \hat{b}^{\dot\mu}\rbrack\lbrack\hat{D}_{\alpha}, \hat{b}^{\dot\rho}\rbrack
-\epsilon_{\alpha\beta}\lbrack\hat{D}_{\dot\rho}, \hat{b}^{\dot\mu}\rbrack\hat{F}^{\beta\dot\rho}
-\epsilon_{\alpha\beta}\lbrack\hat{D}^{\dot\mu}, \hat{b}_{\dot\rho}\rbrack\hat{F}^{\beta\dot\rho}
\nn\\
&&
+\hat{F}^{\dot\mu\dot\rho}\hat{F}_{\alpha\dot\rho}
+\epsilon_{\alpha\beta}\hat{F}^{\dot\mu\dot\rho}\lbrack\hat{D}^{\beta}, \hat{b}_{\dot\rho}\rbrack)
+{\cal O}(\theta^2);
\nn\\
&&
\hat{{\cal F}}_{\alpha\dot\mu}
\nn\\
&=&\hat{F}_{\alpha\dot\rho}
+\theta(-\lbrack\hat{D}_{\dot\mu}, \hat{b}^{\dot\rho}\rbrack \hat{F}_{\alpha\dot\rho}
+\epsilon_{\alpha\beta}\hat{F}_{\dot\mu\dot\rho}\hat{F}^{\beta\dot\rho}
+\hat{F}_{\dot\mu\dot\rho}\lbrack\hat{D}_{\alpha}, \hat{b}^{\dot\rho}\rbrack)
+{\cal O}(\theta^2);
\nn\\
&&
\frac{1}{2\theta}\epsilon^{\alpha\beta}\mathrm{Str}(\hat{{\cal F}}_{\alpha\beta}-\hat{F}_{\alpha\beta})
\nn\\
&=&\epsilon^{\alpha\beta}\mathrm{Str}\bigg\lbrack(-\lbrack\hat{D}_{\beta}, \hat{b}^{\dot{\mu}}\rbrack\hat{F}_{\alpha\dot{\mu}}
-\epsilon_{\beta\gamma}F_{\alpha\dot{\mu}}F^{\gamma\dot{\mu}})
\nn\\
&&
+\theta\bigg(-\frac{3}{2}\hat{F}_{\alpha\dot{\mu}}\hat{F}_{\beta\dot{\sigma}}\hat{F}^{\dot{\mu}\dot{\sigma}}
\nn\\
&&
+2\epsilon_{\beta\gamma}\lbrack \hat{D}^{\dot{\mu}}, \hat{b}_{\dot{\sigma}}\rbrack\hat{F}_{\alpha\dot{\mu}}\hat{F}^{\gamma\dot{\sigma}}
-2\epsilon_{\beta\gamma}\lbrack\hat{D}^{\gamma}, \hat{b}_{\dot{\sigma}}\rbrack\hat{F}_{\alpha\dot{\mu}}\hat{F}^{\dot{\mu}\dot{\sigma}}
\nn\\
&&
+\lbrack\hat{D}^{\dot{\sigma}}, \hat{b}^{\dot{\mu}}\rbrack\lbrack\hat{D}_{\alpha}, \hat{b}_{\dot{\sigma}}\rbrack\hat{F}_{\dot{\mu}\beta}
+\frac{1}{2}\lbrack\hat{D}_{\alpha}, \hat{b}^{\dot{\mu}}\rbrack\lbrack\hat{D}_{\beta}, \hat{b}^{\dot{\nu}}\rbrack\hat{F}_{\dot{\mu}\dot{\nu}}\bigg)
\bigg\rbrack
+{\cal O}(\theta^2).
\eea
We now show other necessary covariant field strengths in the commutative description:
\bea
&&
\hat{{\cal H}}_{\dot1\dot2\dot3}
\nn\\
&=&H_{\dot1\dot2\dot3}
\nn\\
&&
+\theta\bigg((\partial_{\dot\mu}b^{\dot\rho, \mathrm{U(1)}})\lbrack D_{\dot\rho}, b^{\dot\mu, \mathrm{SU(N)}}\rbrack
+b^{\dot\rho, \mathrm{U(1)}}\partial_{\dot\rho}(\lbrack D_{\dot\mu}, b^{\dot\mu, \mathrm{SU(N)}}\rbrack)
\nn\\
&&
+\frac{1}{2}(\partial_{\dot\mu}b^{\dot\rho, \mathrm{U(1)}})(\partial_{\dot\rho}b^{\dot\mu, \mathrm{U(1)}})
+b^{\dot\rho, \mathrm{U(1)}}(\partial_{\dot\mu}\partial_{\dot\rho}b^{\dot\mu, \mathrm{U(1)}})
+\frac{1}{2}(\partial_{\dot\mu}b^{\dot\mu, \mathrm{U(1)}})(\partial_{\dot\rho}b^{\dot\rho, \mathrm{U(1)}})
\bigg)
\nn\\
&&
+\frac{\theta}{2}(\lbrack D_{\dot{\nu}}, b^{\dot{\nu}}\rbrack\lbrack D_{\dot{\rho}}, b^{\dot{\rho}}\rbrack
-\lbrack D_{\dot{\nu}}, b^{\dot{\rho}}\rbrack\lbrack D_{\dot{\rho}}, b^{\dot{\nu}}\rbrack)
+{\cal O}(\theta^2);
\nn\\
&&
\hat{{\cal F}}_{\dot\mu\dot\nu}
\nn\\
&=&F_{\dot\mu\dot\nu}
\nn\\
&&+\theta\big(
(\partial_{\dot\mu}b^{\dot\rho, \mathrm{U(1)}})F_{\dot\rho\dot\nu}
+b^{\dot\rho, \mathrm{U(1)}}(\partial_{\dot\rho}F_{\dot\mu\dot\nu})
+(\partial_{\dot\nu}b^{\dot\rho, \mathrm{U(1)}})F_{\dot\mu\dot\rho}\big)
\nn\\
&&
+\theta(\lbrack D_{\dot{\sigma}}, b^{\dot{\sigma}}\rbrack F_{\dot{\mu}\dot{\nu}}
-\lbrack D_{\dot{\mu}}, b^{\dot{\sigma}}\rbrack F_{\dot{\sigma}\dot{\nu}}
-\lbrack D_{\dot{\nu}}, b^{\dot{\sigma}}\rbrack F_{\dot{\mu}\dot{\sigma}})
+{\cal O}(\theta^2);
\nn\\
&&
\hat{{\cal F}}_{\alpha\dot\mu}
\nn\\
&=&F_{\alpha\dot\mu}
\nn\\
&&
+\theta\big(
(\partial_{\alpha}b^{\dot\rho, \mathrm{U(1)}})F_{\dot\rho\dot\mu}
+b^{\dot\rho, \mathrm{U(1)}}(\partial_{\dot\rho}F_{\alpha\dot\mu})
+(\partial_{\dot\mu}b^{\dot\rho, \mathrm{U(1)}})F_{\alpha\dot\rho}\big)
\nn\\
&&
+\theta(-\lbrack D_{\dot\mu}, b^{\dot\rho}\rbrack F_{\alpha\dot\rho}
+\epsilon_{\alpha\beta}F_{\dot\mu\dot\rho}F^{\beta\dot\rho}
+F_{\dot\mu\dot\rho}\lbrack D_{\alpha}, b^{\dot\rho}\rbrack)
+{\cal O}(\theta^2).
\eea
The last covariant object in the commutative description that we need is
\bea
&&
\int d^2xd^3\dot{y}\ \mathrm{Str}\bigg(\frac{1}{2\theta}\epsilon^{\alpha\beta}\hat{{\cal F}}_{\alpha\beta}\bigg)
\nn\\
&\sim&\mathrm{Str}\bigg\lbrack
(-F^{\mathrm{SU(N)}}_{\alpha\dot{\mu}}\lbrack D_{\beta}, b^{\dot{\mu}, \mathrm{SU(N)}}\rbrack
-\epsilon_{\beta\gamma}F_{\alpha\dot{\mu}}^{\mathrm{SU(N)}}F^{\gamma\dot{\mu}, \mathrm{SU(N)}})
\nn\\
&&
+\theta\bigg(-3\epsilon^{\alpha\beta}F^{\mathrm{U(1)}}_{\alpha\dot{\mu}}F^{\mathrm{SU(N)}}_{\beta\dot{\sigma}}F^{\dot{\mu}\dot{\sigma}, \mathrm{SU(N)}}
-\frac{3}{2}\epsilon^{\alpha\beta}F^{\mathrm{SU(N)}}_{\alpha\dot{\mu}}F^{\mathrm{SU(N)}}_{\beta\dot{\sigma}}F^{\dot{\mu}\dot{\sigma}, \mathrm{U(1)}}
\nn\\
&&
+2\lbrack D^{\dot{\mu}}, b^{\mathrm{SU(N)}}_{\dot{\sigma}}\rbrack F^{\mathrm{SU(N)}}_{\alpha\dot{\mu}}F^{\alpha\dot{\sigma}, \mathrm{U(1)}}
+2\lbrack  D^{\dot{\mu}}, b^{\mathrm{SU(N)}}_{\dot{\sigma}}\rbrack F^{\mathrm{U(1)}}_{\alpha\dot{\mu}}F^{\alpha\dot{\sigma}, \mathrm{SU(N)}}
\nn\\
&&
-2\lbrack D^{\alpha}, b^{\mathrm{SU(N)}}_{\dot{\sigma}}\rbrack F_{\alpha\dot{\mu}, \mathrm{U(1)}}F^{\dot{\mu}\dot{\sigma}, \mathrm{SU(N)}}
-2\lbrack D^{\alpha}, b^{\mathrm{SU(N)}}_{\dot{\sigma}}\rbrack F_{\alpha\dot{\mu}, \mathrm{SU(N)}}F^{\dot{\mu}\dot{\sigma}, \mathrm{U(1)}}
\nn\\
&&
+H_{\dot1\dot2\dot3}^{\mathrm{U(1)}}F^{\mathrm{SU(N)}}_{\alpha\dot\mu}F^{\alpha\dot\mu, \mathrm{SU(N)}}
\nn\\
&&
+\epsilon^{\alpha\beta}\lbrack D^{\dot\sigma}, b^{\dot\mu, \mathrm{SU(N)}}\rbrack\lbrack D_{\alpha}, b^{\mathrm{SU(N)}}_{\dot\sigma}\rbrack F_{\dot\mu\beta}^{\mathrm{U(1)}}
+\epsilon^{\alpha\beta}\lbrack D_{\alpha}, b^{\dot\mu, \mathrm{SU(N)}}\rbrack H_{\dot1\dot2\dot3}^{\mathrm{U(1)}} F_{\dot\mu\beta}^{\mathrm{SU(N)}}\bigg)\bigg\rbrack
+{\cal O}(\theta^2),
\nn\\
\eea
in which we ignore the pure U(1) terms that was already computed in Ref. \cite{Ma:2020msx}.

\subsection{Lagrangian}
\noindent
The result of commutative description for the pure U(1) sector is \cite{Ho:2011yr}
\bea
&&S_{\mathrm{RR41}}
\nn\\
&=&\frac{T_4}{\theta^2}\int d^2xd^3\dot{y}\ \mathrm{Str}\bigg\lbrack\bigg(-\frac{1}{4}F_{\dot\mu\dot\nu}^{\mathrm{U(1)}}F^{\dot\mu\dot\nu, \mathrm{U(1)}}
-\frac{1}{2}F_{\alpha\dot\nu}^{\mathrm{U(1)}}F^{\alpha\dot\nu, \mathrm{U(1)}}
-\frac{1}{4}F_{\alpha\beta}^{\mathrm{U(1)}}F^{\alpha\beta, \mathrm{U(1)}}
\nn\\
&&+\theta\bigg(
\frac{1}{2}F_{01}^{\mathrm{U(1)}}F_{01}^{\mathrm{U(1)}}F_{01}^{\mathrm{U(1)}}+\frac{1}{4}F_{01}^{\mathrm{U(1)}}F^{\dot{\mu}\dot{\nu}, \mathrm{U(1)}}F_{\dot{\mu}\dot{\nu}}^{\mathrm{U(1)}}
\nn\\
&&
-\frac{1}{2}F_{01}^{\mathrm{U(1)}}F^{\alpha\dot{\mu}, \mathrm{U(1)}}F_{\alpha\dot{\mu}}^{\mathrm{U(1)}}
-\frac{1}{2}\epsilon_{\alpha\beta}F^{\alpha\dot{\mu}, \mathrm{U(1)}}F^{\beta\dot{\nu}, \mathrm{U(1)}}F_{\dot{\mu}\dot{\nu}}^{\mathrm{U(1)}}
\bigg)\bigg\rbrack
+{\cal O}(\theta^2)
.
\eea
In the non-commutative R-R D4-brane, the non-local derivative appears in the first-order correction.
However, it has been removed from the commutative description \cite{Ma:2020msx}.
Hence, it implies that the non-locality is not inherent \cite{Ma:2020msx}.
\\

\noindent
For the non-Abelian sectors, we can use a similar way to integrate out the non-dynamical field $b^{\dot\mu, \mathrm{SU(N)}}$, equivalent to using
\bea
H_{\dot1\dot2\dot3}^{\mathrm{SU(N)}}=-F_{01}^{\mathrm{SU(N)}},
\eea
and then rewrite the gauge potential as
\bea
b^{\dot\mu, \mathrm{SU(N)}}=D^{\dot\mu}\nabla^{-1}H^{\mathrm{SU(N)}}_{\dot1\dot2\dot3}
=-D^{\dot\mu}\nabla^{-1}F^{\mathrm{SU(N)}}_{01},
\eea
where
\bea
\nabla\equiv D_{\dot\rho}D^{\dot\rho}.
\eea
The resulting Lagrangian for the non-Abelian sector is given by
\bea
&&
S_{\mathrm{RR41N}}
\nn\\
&=&\frac{T_4}{\theta^2}\int d^2xd^3\dot{y}\
\mathrm{Str}\bigg\lbrack\bigg(-\frac{1}{4}F_{\dot\mu\dot\nu}^{\mathrm{SU(N)}}F^{\dot\mu\dot\nu, \mathrm{SU(N)}}
-\frac{1}{2}F_{\alpha\dot\nu}^{\mathrm{SU(N)}}F^{\alpha\dot\nu, \mathrm{SU(N)}}
-\frac{1}{4}F_{\alpha\beta}^{\mathrm{SU(N)}}F^{\alpha\beta, \mathrm{SU(N)}}\bigg)
\nn\\
&&
+\theta\bigg(\frac{1}{2}F_{01}^{\mathrm{U(1)}}F_{01}^{\mathrm{SU(N)}}F_{01}^{\mathrm{SU(N)}}
-(D^{\dot\rho}\nabla^{-1}F_{01}^{\mathrm{SU(N)}})F_{01}^{\mathrm{SU(N)}}(\partial_{\dot\rho}F_{01}^{\mathrm{U(1)}})
\nn\\
&&
+\frac{1}{4}F_{01}^{\mathrm{U(1)}}F_{\dot\mu\dot\nu}^{\mathrm{SU(N)}}F^{\dot\mu\dot\nu, \mathrm{SU(N)}}
-\frac{1}{2}F_{01}^{\mathrm{U(1)}}F_{\alpha\dot\mu}^{\mathrm{SU(N)}}F^{\alpha\dot\mu, \mathrm{SU(N)}}
-H_{\dot1\dot2\dot3}^{\mathrm{SU(N)}}F_{\dot\mu\dot\nu}^{\mathrm{SU(N)}}F^{\dot\mu\dot\nu, \mathrm{U(1)}}
\nn\\
&&
-\frac{1}{2}\epsilon^{\alpha\beta}F^{\mathrm{SU(N)}}_{\alpha\dot{\mu}}F^{\mathrm{SU(N)}}_{\beta\dot{\sigma}}F^{\dot{\mu}\dot{\sigma}, \mathrm{U(1)}}
+3\epsilon^{\alpha\beta}F^{\mathrm{U(1)}}_{\alpha\dot{\mu}}F^{\mathrm{SU(N)}}_{\beta\dot{\sigma}}F^{\dot{\mu}\dot{\sigma}, \mathrm{SU(N)}}
\nn\\
&&
+\epsilon^{\alpha\beta}\lbrack D^{\dot\sigma}, D^{\dot\mu}\nabla^{-1}F_{01}^{\mathrm{SU(N)}}\rbrack\lbrack D_{\alpha}, D_{\dot\sigma}\nabla^{-1}F_{01}^{\mathrm{SU(N)}}\rbrack F_{\dot\mu\beta}^{\mathrm{U(1)}}
\nn\\
&&
+\epsilon^{\alpha\beta}\lbrack D_{\alpha}, D^{\dot\mu}\nabla^{-1}F_{01}^{\mathrm{SU(N)}}\rbrack F_{01}^{\mathrm{U(1)}} F_{\dot\mu\beta}^{\mathrm{SU(N)}}
\nn\\
&&
-\lbrack D_{\dot\mu}, D^{\dot\sigma}\nabla^{-1}F_{01}^{\mathrm{SU(N)}}\rbrack F_{\dot\sigma\dot\nu}^{\mathrm{U(1)}}F^{\dot\mu\dot\nu, \mathrm{SU(N)}}
-\lbrack D_{\dot\nu}, D^{\dot\sigma}\nabla^{-1}F_{01}^{\mathrm{SU(N)}}\rbrack F_{\dot\mu\dot\sigma}^{\mathrm{SU(N)}}F^{\dot\mu\dot\nu, \mathrm{U(1)}}
\nn\\
&&
-\lbrack D_{\dot\mu}, D^{\dot\rho}\nabla^{-1}F_{01}^{\mathrm{SU(N)}}\rbrack F_{\alpha\dot\rho}^{\mathrm{SU(N)}}F^{\alpha\dot\mu, \mathrm{U(1)}}
-\lbrack D_{\dot\mu}, D^{\dot\rho}\nabla^{-1}F_{01}^{\mathrm{SU(N)}}\rbrack F_{\alpha\dot\rho}^{\mathrm{U(1)}}F^{\alpha\dot\mu, \mathrm{SU(N)}}
\nn\\
&&
+\lbrack D_{\alpha}, D^{\dot\rho}\nabla^{-1}F_{01}^{\mathrm{SU(N)}}\rbrack F_{\dot\mu\dot\rho}^{\mathrm{SU(N)}}F^{\alpha\dot\mu, \mathrm{U(1)}}
+\lbrack D_{\alpha}, D^{\dot\rho}\nabla^{-1}F_{01}^{\mathrm{SU(N)}}\rbrack F_{\dot\mu\dot\rho}^{\mathrm{U(1)}}F^{\alpha\dot\mu, \mathrm{SU(N)}}
\bigg)\bigg\rbrack
\nn\\
&&
+{\cal O}(\theta^2).
\eea
The non-local derivative cannot be removed in the commutative description, unlike in the Abelian sector.
Hence, it implies that the non-locality is genuine, which makes the difference between the Abelian and non-Abelian sectors.
The genuine non-locality is due to the covariant gauge potential $b^{\dot\mu}$, which only appears in the SU($N$) and is lost in the U(1) part.
Since we only work to first order, the result is generic to R-R D-branes, not just D4-branes.

\subsection{Implication}
\noindent
Let us now discuss the implications of the commutative description for the form of DBI and the M5-branes.
The most naive approach is to perform the electromagnetic dual of the 4D DBI in the commutative description and then replace the U(1) gauge group with U($N$).
In the NS-NS D-brane, one can only replace the gauge group to promote it to multiple D-branes in both commutative and non-commutative settings.
However, this situation does not continue to the R-R D-branes.
In the non-commutative R-R D-branes, we can combine the U(1) and SU($N$) fields into a U($N$) field.
However, the combination does not work in the commutative theory.
Hence, we should perform the electromagnetic dual from the non-commutative DBI theory and then replace the gauge group.
However, it has a technical difficulty.
If we perform the electromagnetic dual from the commutative DBI theory and then apply the SW map to show the non-commutative DBI theory, and then replace the gauge group, it is also hard to see the elegant form of the DBI theory.
\\

\noindent
For the M5-branes, we expect the non-local terms to appear in the non-Abelian sector.
This enables the formulation of multiple M5-branes to differ from that of a single M5-brane.
To obtain the R-R D4-branes from compactification, it is necessary to introduce 5D boundary gauge fields, along with the coupling of the 6D self-dual gauge fields \cite{Chu:2011fd}.
The 5D one-form gauge fields are the function of the 6D self-dual gauge fields \cite{Chu:2011fd}.
With the higher-derivative correction, the 5D gauge field should generate the non-locality similar to the non-Abelian sector of the R-R D-branes.
However, the Lagrangian formulation for the non-commutative R-R D4-branes depends only on the covariant field strengths \cite{Ma:2023hgi}; we expect that the non-commutative M5-branes should be easier to construct in this setting.
 
\section{Discussion and Conclusion}
\label{sec:5}
\noindent
 In this work, we have investigated the commutative description of D-branes in a large R–R field background using the Seiberg–Witten (SW) map. We extended the SW-map solution from the U(1) case \cite{Ma:2020msx} to the U($N$) setting for flat backgrounds. 
A key result is that, unlike in the non-commutative formulation, the commutative description cannot be written purely in terms of U($N$) fields. 
This stands in sharp contrast with NS–NS D-branes, where both commutative and non-commutative descriptions can be consistently expressed using U($N$) gauge fields \cite{Seiberg:1999vs}. 
Consequently, the gauge group in the commutative R–R D-brane theory cannot be naively promoted from U(1) to U($N$).
\\

\noindent
Our analysis shows that the origin of this discrepancy lies in the appearance of the covariant gauge potential in the non-Abelian sector, which introduces genuine non-locality into the commutative description of R–R D-branes. 
The distinction between Abelian and non-Abelian sectors is reflected in their transformation properties. 
In the Abelian case, the non-local operators present in the non-commutative theory \cite{Ho:2011yr} can be eliminated upon passing to the commutative description \cite{Ma:2020msx}. 
This feature explains why, in the non-Abelian sector, five-dimensional (5D) gauge fields must couple to the six-dimensional (6D) self-dual gauge fields. 
In contrast, such coupling is unnecessary in the Abelian sector.
\\

\noindent
 Our explicit SW-map solution is obtained to first order in the non-commutativity parameter, without introducing ambiguity. 
Nevertheless, it should be possible to extend the solution to all orders, allowing for a systematic study of how non-locality manifests in M5-branes from the perspective of R–R D4-branes. 
An auspicious direction is the construction of non-commutative M5-brane theories that combine the NP M5-brane formulation \cite{Ho:2008nn,Ho:2008ve} with the introduction of a 5D nondynamical one-form gauge field coupled to the 6D self-dual field strength \cite{Chu:2011fd}. 
Understanding precisely how the 5D gauge fields generate non-locality would offer valuable insight into the structure of multiple M5-branes.

\section*{Acknowledgments}
\noindent
The author would like to thank thank Nan-Peng Ma for his encouragement.



\end{document}